\documentclass[preprint,aps,prb,showpacs,amsmath,amssymb]{revtex4}
\usepackage{graphics}
\usepackage{epsf,graphicx,pstcol,epsfig,psfrag}
\begin{document}

\title{Frustration - no frustration  crossover and phase transitions  in 2D spin models with zig-zag structure}

\author{Jozef Sznajd}
\address{Institute for Low Temperature and Structure Research, Polish Academy of Sciences, 50-422 Wrocław, Poland.}

\begin{abstract}
Three 2D spin models made of frustrated zig-zag chains with competing interactions which by exact summation with respect to some degrees of freedom can be replaced by an effective temperature-dependent interaction  were considered.
The first model, exactly solvable Ising chains coupled by only four-spin interactions does not exhibit any finite temperature phase transition, nevertheless temperature can trigger a frustration - no frustration crossover accompanied by gigantic specific heat. A similar effect was observed in several two-leg ladder models [Yin Weiguo, arXiv:2006.08921v2 (2020), 2006.15087v1 (2020)]. The anisotropic Ising chains coupled by a direct interchain interaction and competing with it indirect interaction via spins located between chains are analyzed by using exact Onsager's equation and linear perturbation renormalization  group (LPRG). Depending on the parameter set such a model exhibits one antiferromagnetic (AF) or ferromagnetic (FM) phase transition or three phase transitions with reentrant disordered phase between AF and FM ones. The LPRG method was also used to study coupled uniaxial $XXZ$ chains which, for example, can be a minimal model to describe the magnetic properties of compounds in which uranium and rare earth atoms form zig-zag chains.  As with the Ising model for a certain set of parameters the model can undergo three phase transitions. However, both inchain and interchain plain interactions $s_{i,j}^x s_{k,l}^x+s_{i,j}^y s_{k,l}^y$ can eliminate the reentrant disordered phase and then only one transition takes place. Additionally, the $XXZ$ model can undergo  temperature-induced metamagnetic transition.
\end{abstract}
\maketitle
 
\section{Introduction}

In statistical physics a phase transition is rigorously defined as a point of non-analyticity of free energy which is only possible in the thermodynamic limit. However, in practice in experimental studies or computer simulations, the phase transition point is often taken to be the "anomaly" point of the response function on either side of which the states of the system can be distinguished.  Such a definition is most commonly accepted outside physics ( e.g., in sociophysics \cite{Toral}) but also sometimes in condensed matter. Thus, it is useful on the one hand to have a tool that allows to distinguish between a crossover and phase transition in models that are not exactly solvable (e.g. attempt propose in Ref.~\onlinecite{Hu}) and, on the other hand, understand mechanisms that can lead to the peculiarity of the response function despite the absence of a phase transition.
Such peculiarities, namely the entropy jump and a  gigantic peak in specific heat, have  recently been observed in several decorated one dimensional spin models with short range interactions and frustration  \cite{Str, Roj1, Roj2, Roj3, Yin1, Hutak, Yin2, Roj4} at low but finite temperatures. 
A paradigm is a two-leg spin ladder  \cite{Yin1} with $s_i$ spins in the legs and with an additional frustrating rung  spins $b_i$.
This behavior resembles a phase transition, contrary to the strict theory which clearly indicates that in the one-dimensional models considered in the above papers, a phase transition is not possible.
In these decorated  models after removing redundant spins $b_i$ an effective interaction between remaining spins $s_i$ depends on temperature and can vanish at a certain temperature \cite{Yin1} $T = T^*$ as a result of the competition of direct interaction between $s_i$ spins and indirect via $b_i$ spins. Consequently, the system may be frustrated for temperatures on one side of the $T^*$ and unfrustrated on the other side. Thus, it can be assumed that the decisive reason for the peculiar behavior of thermodynamic quantities near $T^*$ is frustration \cite{Yin1}. Studying the two-leg ladder Ising model with trimer rungs Hutak at al \cite{Hutak} stated 
that to the emergence of anomalous behavior in such a model "frustration is not vitally necessary; the only demand is to have a suitable $J_{\bot}(T)$" -
effective, temperature-dependent interaction which vanishes at $T=T^*$ and then "the effective model reduces to the standard Ising-chain". So the peculiar low-temperature behaviors  of the considered model "are related to the criticality of the standard Ising chain at zero temperature". However, to our knowledge the frustration  has been present in all one-dimensional models exhibiting singular behavior of thermodynamic quantities despite the absence of a phase transition, hereinafter referred to as the crossover. Moreover, it seems that the existence of the temperature $T^*$ at which the effective interaction parameter changes 
sign is a necessary but not sufficient condition for the crossover. 
The singularity emerges when $T^*$ coincides with the temperature of bounding between some subsystems of spins accompanied by maximum of the specific heat and inflection point of the spin-spin nearest neighbor (NN) correlation function \cite{JS3}.

     The purpose of this paper is to study the effect of the temperature-dependent effective interaction discussed above on the thermodynamic properties of the two-dimensional Ising and $XXZ$ models.
To do so we consider systems of the coupling chains with zig-zag structure. First in Sec.II, the exactly solvable  Ising model with only four-spin coupling which by a simple transformation can be formally reduced to a one dimensional system \cite{JS1}, nevertheless the four-spin interaction is crucial for crossover to occur. Such a system does not, of course, exhibit a phase transition. 
The phase transitions are analyzed in Sec.III in the systems of Ising chains with zig-zag interaction and  standard two-spin coupling and in Sec. IV, in $XXZ$ zig-zag chains with easy axis. Our motivation to consider this model comes also from studies indicating that taking  into account the presence of zig-zag chains of uranium or rare earth atoms in crystal structures of several magnetic compounds,  seems to be crucial for explaining the character of the magnetic phase transitions \cite{Aoki, Mas, Poole, Pikul}. 

\section{Giant specific heat. Exact results.}
Let us consider  2D Ising model made of spin chains with zig-zag structure interacting via four spin interactions (Fig.1).  The model corresponds to a Hamiltonian 

\begin{equation}
H = -J_y \sum_{i,j} s_{i,j} s_{i+1,j} - j_1\sum_i (s_{i,j}+s_{i+1,j}) b_{i,j} - j_4 \sum_{i,j} s_{i,j} s_{i+1,j}  s_{i,j+1} s_{i+1,j+1},
\end{equation}
where $s_i$ (full circles in Fig.1) and $b_i$ (open circles) denote spins $\pm 1$, the label $i$ refers to rows and $j$ to columns.  The interaction of the spins $s_i$ in a chain is $J_y$ and "zig-zag" interaction between spins $s_i$ and $b_i$ is $j_1$. The chains interact only via four-spin coupling $j_4$.   Because there is not interaction between $b_i$ spins one can rigorously performed summation over $b_i$ spins degrees of freedom \cite{Yin1, Hutak} which leads to an effective Hamiltonian in the form

\begin{equation}
H = - \frac{1}{\beta} N A - j_t \sum_{i,j} s_{i,j} s_{i+1,j}  - j_4 \sum_{i,j} s_{i,j} s_{i+1,j}  s_{i,j+1} s_{i+1,j+1}
\end{equation}
where $j_t = J_y + \frac{1}{\beta} B $ effective interaction between $s_i$ spins in a chain and \cite{Yin1}
\begin{equation}
A = \frac{1}{2} \ln( 4\cosh 2 \beta j_1)   \quad B =  \frac{1}{2} \ln( \cosh 2 \beta  j_1).
\end{equation}

The free energy of the model described by the Hamiltonian (2) can be easily found exactly by the simple transformation $s_i s_{i+1} = \Omega_i$
\begin{equation}
H = - \frac{1}{\beta} N A -j_t \sum_{i,j} \Omega_{i,j} - j_4 \sum_{i,j} \Omega_{i,j} \Omega_{i,j+1}.
\end{equation}
which is in fact one dimensional model with an exact free energy per spin \cite{JS1} $F = -f/\beta$.
\begin{equation}
f = A + \ln[\exp(\beta j_4) \cosh (\beta j_t) + \sqrt{ \exp(2 \beta j_4) \sinh^2 (\beta j_t) + \exp(-2 \beta j_4)}
\end{equation}
As shown in Fig.2 for the original antiferromagnetic interaction $J_y = -1$ the Hamiltonian (4) describes three possible  ground state phases defined by the NN correlation functions:
\begin{equation}
G_{ss} \equiv \langle
\Omega_{i,j} \rangle \equiv \langle s_{i,j} s_{i+1,j} \rangle = \frac{\partial f}{\partial J_y}  , \quad\mbox{and}  \quad  G_{sb} \equiv \langle s_{i,j} b_{i,j} \rangle =\frac{\partial f}{\partial j_1}.
\end{equation}

\begin{figure}
\label{Fig_1}
 \epsfxsize=12cm \epsfbox{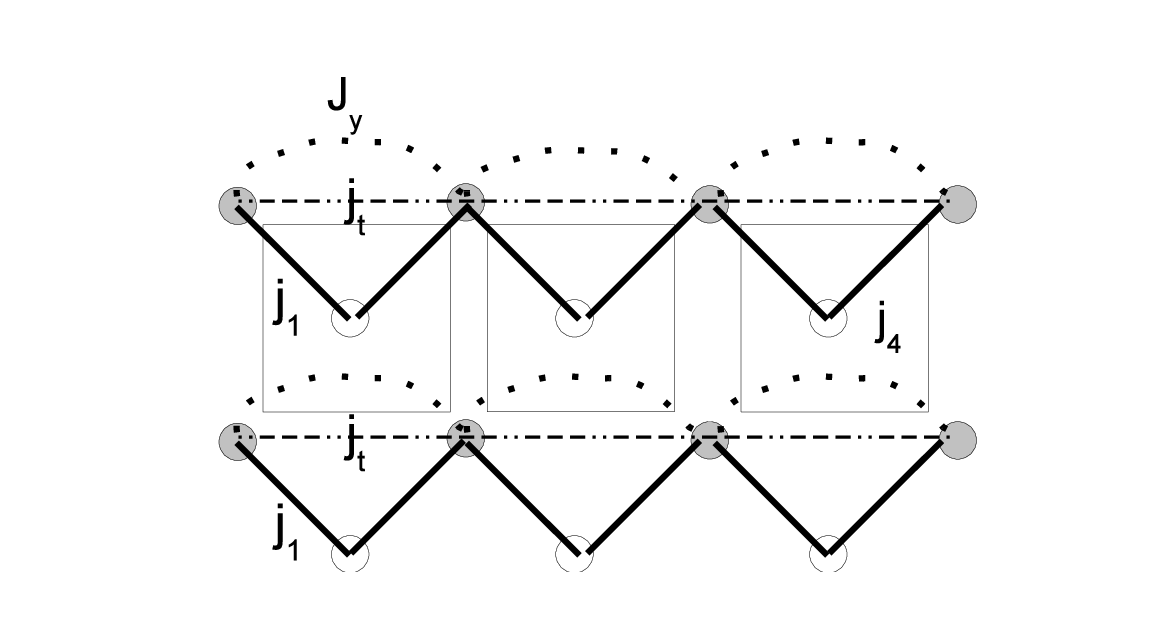}
 \caption{Ising model with  inchain $J_y$, interchain four-spin $j_4$ couplings and zig-zag structure. Full and open circles denote Ising spins $\pm 1$ $s_i$ and $b_i$, respectively.
$j_t$ indicates an effective inchain interaction after summing over $b_i$ spins. }
 \end{figure}
As seen in the phases $AF$ ($G_{ss} = -1, G_{sb}=0$) and FM ($G_{ss} = 1, G_{sb}=1$) the long range (LO) ground state order in individual chain is antiferromagnetic with extinction of the moments $b_{i,j}$ (AF) due to frustration, and fully ferromagnetic (FM), respectively. In  the third phase alternate chains are ferro- and antiferromagnetic. A positive four-spin interaction $j_4>0$ does not affect the ground state, not coupling individual chains, while negative can trigger zero-temperature phase transition from $AF$ or $FM$ phases to $M$ with correlations on the borderline $AF-M$, $G_{ss}=-1/\sqrt{5}, G_{sb} = (1-1/\sqrt{5})/2$  and on the borderline $FM-M$, $G_{ss}=1/\sqrt{5}, G_{sb} = (1+1/\sqrt{5})/2$ (Fig.2) \cite{Yin1, Yin3, Red}. 
Of course the temperature destroys the LO and there is no finite temperature phase transition in the considered model. Nevertheless for $j_4 >0$  the specific heat shows peculiar behavior - huge increase at some finite temperature, which resembles singularity at the critical point, as with the aforementioned spin ladders.
\begin{figure}
\label{Fig_2}
 \epsfxsize=10cm \epsfbox{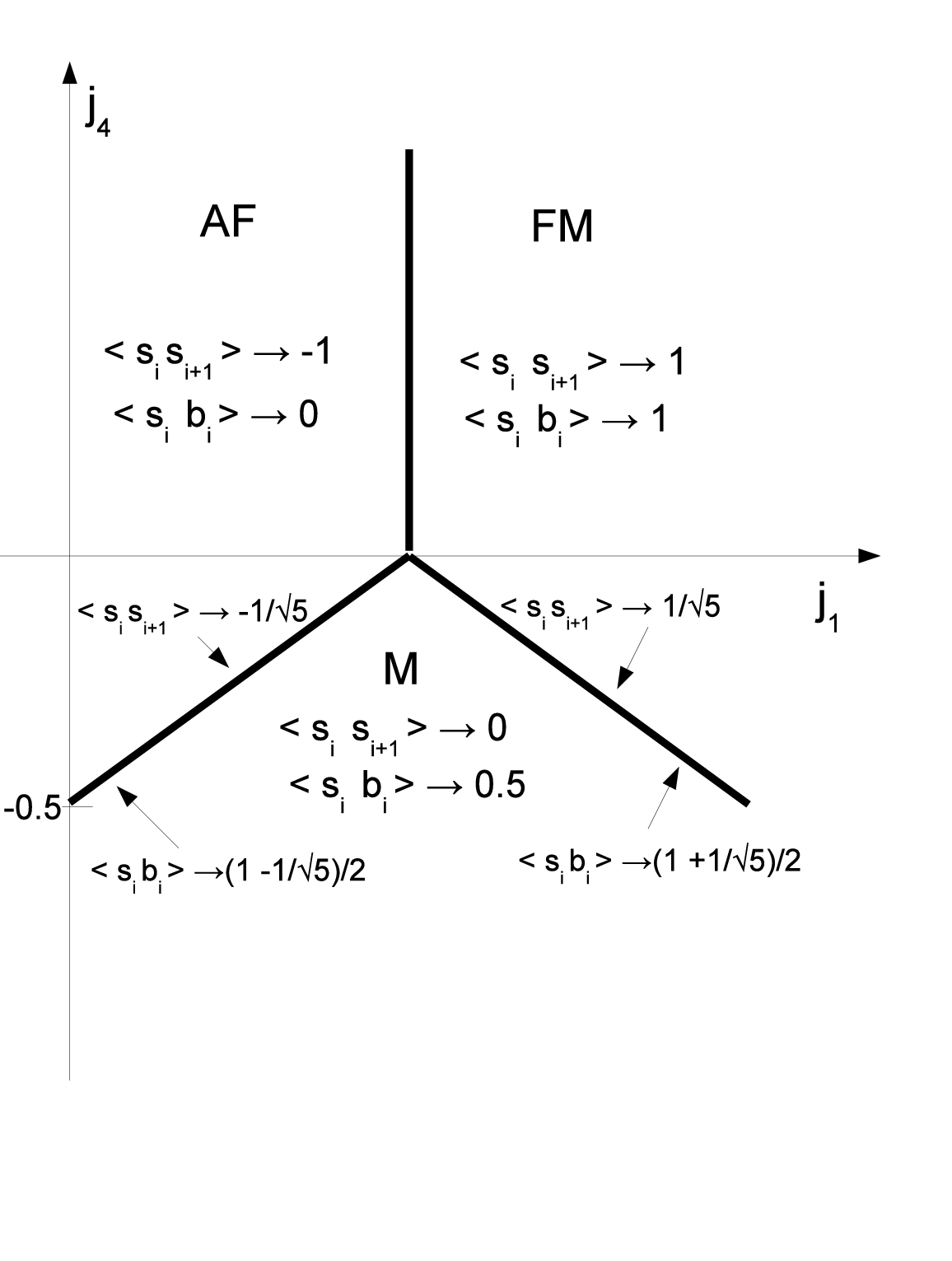}
 \caption{Zero temperature phase diagram of the 2D Ising model in the plane ($j_1$, $j_4$) (1) with inchain interaction $J_y = -1$ and zig-zag structure (Fig.1)}
 \end{figure}

\begin{figure}
\label{Fig_3}
 \epsfxsize=16.5cm \epsfbox{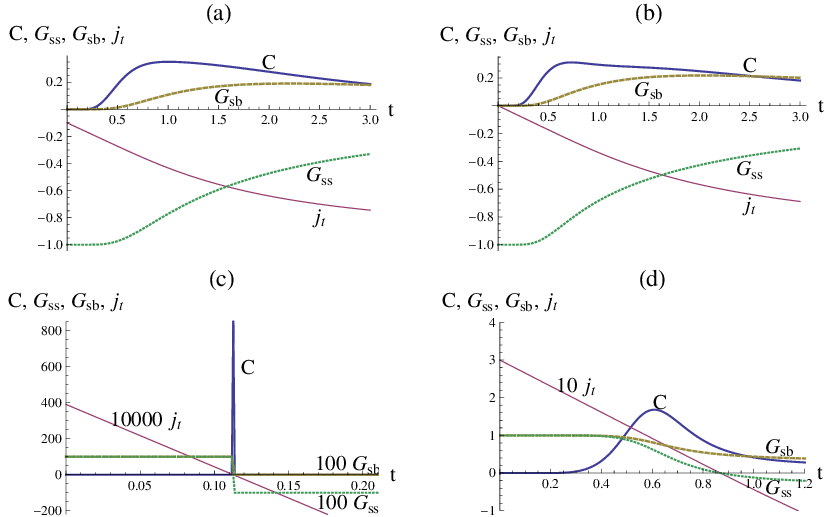}
 \caption{Temperature dependences of the specific heat (solid bold lines), NN correlation functions $G_{ss}$ (dotted), $G_{sb}$ (dashed) and effective interaction $j_t$ (thin) for $J_y = -1, j_4 = 0.5$, and several values of $j_1$: 0 (a), 0.9 (b), 1.039076 (c), and 1.3 (d).}
 \end{figure}
 
  \begin{figure}
\label{Fig_4}
 \epsfxsize=16.5cm \epsfbox{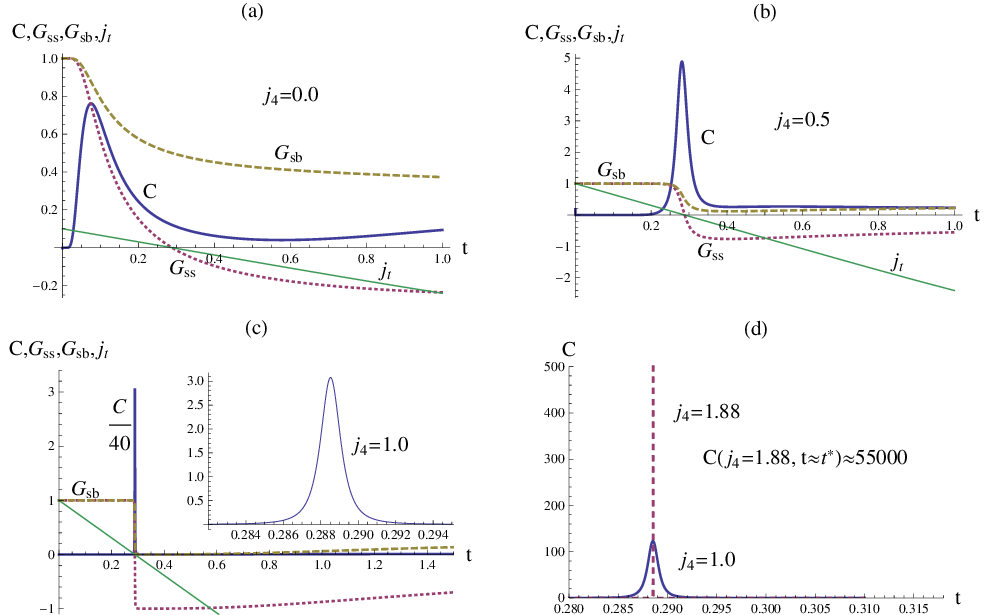}
 \caption{Temperature dependences of the specific heat (solid bold lines), correlation functions $G_{ss}$ (dotted), $G_{sb}$ (dashed) and effective interaction $j_t$ (thin) for $J_y = -1, j_1 = 1.1$, and several values of $j_4$: 0, 0.5, 1.0 and 1.88.  For comparison in the bottom right plot (d) the specific heat for two cases $j_4=1$ and $j_4=1.88 \approx j_4^*$  is presented.}
 \end{figure}

Differentiating the free energy (5) with respect to temperature $t$, $J_y$, and $j_1$ we obtain the specific heat $C$ and correlation functions $G_{ss}$ and $G_{sb}$, respectively. The temperature dependences of these quantities for $j_4 = 0.5$ and several values of $j_1$ are shown in Fig.3, and for fixed value of $j_1=1.1$ and several values of $j_4$ in Fig.4. We should emphasize that all formulas used for the thermodynamic quantities are exact, however, the estimates of the values given e.g. in Fig.4 are numerical and approximate.
Thin lines represent the effective interaction $j_t$. We always assume original $s_{i,j} s_{i+1,j}$ antiferromagnetic  interaction $J_y = -1$.
As seen for $j_1 < -J_y$ (Fig.3a and 3b) the effective  $j_t$ interaction is negative over the entire temperature range and specific heat shows the typical one-dimensional maximum. The correlation function $G_{ss}$ tends to $-1$ and $G_{sb}$ due to frustration to zero, as expected.
For $j_1 > -J_y$ (Fig.3c and 3d) the effective interaction parameter changes sign for some temperature $t=t_p$  accompanied always by a change in the sign of the correlation $G_{ss}  = \langle s_{i,j} s_{i+1,j} \rangle$, which is a necessary condition for a giant jump of the specific heat. However as seen in Fig.3(d) it is not the sufficient one.

\begin{figure}
\label{Fig_5}
 \epsfxsize=10cm \epsfbox{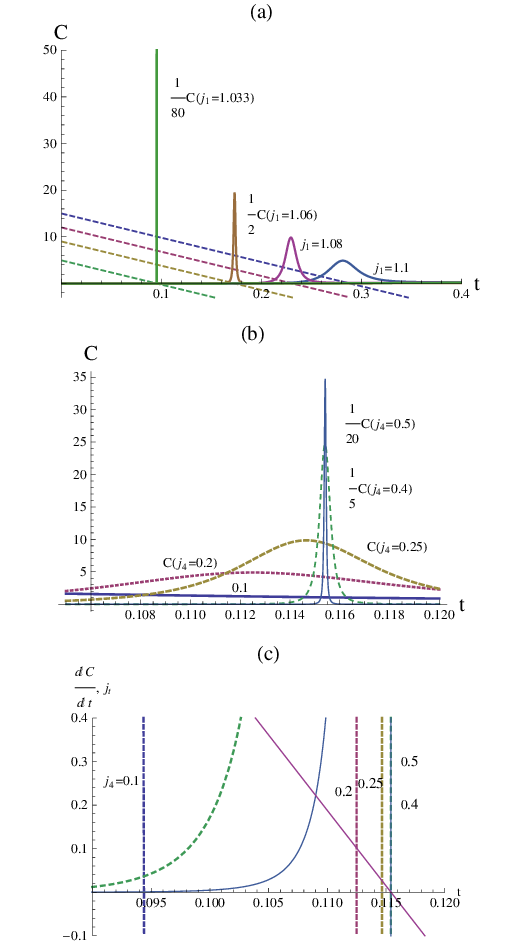}
 \caption{(a) The temperature dependence of the specific heat $C$ for $j_4=0.5$ and several values of $j_1$, dashed lines denote the effective interaction $j_t$ for $j_1=1.1, 1.08, 1.06$ and $1.033$ from top to the bottom, (b) $C$ for $j_1=1.04$ and several values of $j_4$ and  (c) specific heat derivatives $\frac{dC}{dt}$  for $j_1 =1 .04$ and $j_4 = 0.1, 0.2, 0.25, 0.4$ and $0.5$. The intersections of the vertical lines with the abscissa axis mark the positions of the specific heat maxima. Solid thin line indicates the effective interaction $j_t$.}
 \end{figure}

\begin{figure}
\label{Fig_6}
 \epsfxsize=10cm \epsfbox{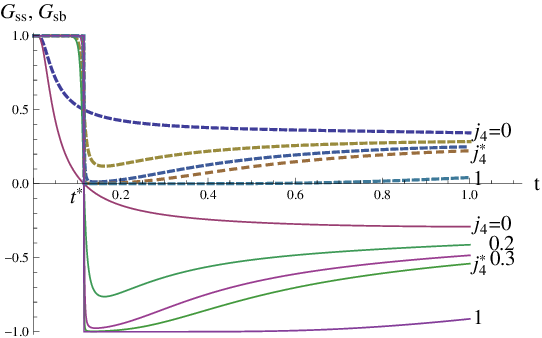}
 \caption{The temperature dependence of the correlation functions $G_{ss}$ (solid lines) and  $G_{sb}$ (dashed), $t^{*} \approx  0.115415603$}
 \end{figure}

\begin{figure}
\label{Fig_7}
 \epsfxsize=13cm \epsfbox{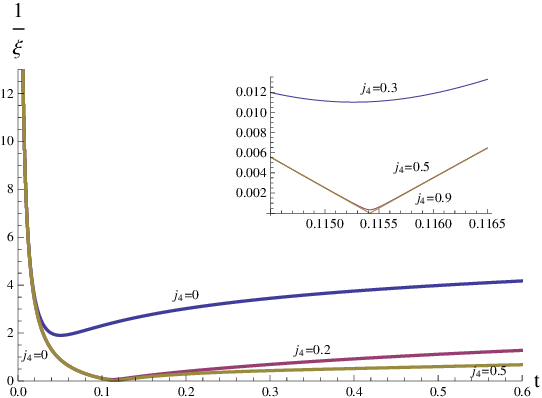}
 \caption{The temperature dependence of the inverse correlation length $1/\xi$  for $j_1 =1 .04$ and $j_4 = 0.0, 0.2$ and $0.5$. The inset shows the inverse correlation length near the temperature of the peculiarity for $j_4 = 0.3, 0.5$ and $0.9$}
 \end{figure}
 
 The crossover accompanied by a gigantic increase in specific heat is not a phase transition and has no precise thermodynamic definition. Nevertheless, in a previous paper \cite{JS3}, we proposed defining such a point in the case of a two-leg spin ladder with triple rungs as the point where the temperature at  which the effective interleg interaction vanishes coincides with the temperature of bounding spins between the ladder legs. Such a definition for the case under consideration is described in Fig.5. In Fig.5a and 5b the temperature dependences of the specific heat $C$ for $j_4=0.5$ and $j_1=1.1,1.08,1.06, 1.033$ and for $j_1=1.04$ and $j_4=0.2,0.25,0.4,05$ are presented, respectively, and in Fig. 5c the specific heat  derivatives $\frac{dC}{dt}$ for $j_1=1.04$ and several values of $j_4$ are shown. As seen in Fig.4 the zig-zag chain with $j_4=0$ shows two maxima instead of the standard one-dimensional one. The low temperature maximum marks the temperature at which the spins $s_i$, $s_j$ become unbound which we denote as $t=t_{max}$. With increasing $j_4$ this maximum increases and moves towards higher temperatures (Fig.5b) up to a certain value of $j_4$. A further increase in $j_4$ changes only infinitesimally the temperature of the position of the specific  heat maximum.  In Fig.5c the temperature $t_{max}(j_4)$ is indicated by the zero of the specific heat derivative  $\frac{dC}{dt}$. As seen this temperature with increase of $j_4$ tends to the temperature at which the effective interaction parameter $j_t$ changes sign $t=t_p$. Fig.5a shows the temperature dependences of the effective interaction $j_t$ (dashed lines) and specific heat (solid lines) for several values of $j_1$. As can be seen, changing the sign of the effective interaction $j_t$ is not a sufficient condition for gigantic heat to occur. Thus, it is possible to take as the characteristic crossover point such a value of $j_1=j_1^{*}$ for given value of  $j_4$ for which $t_{max}=t_p = t^{*}$.  
So, to define a characteristic temperature of crossover we propose $t=t^*$  which can be found from the set of equations $j_t =0$ and $dC/dt =0$. This is shown in Fig.5c where the vertical lines mark the points of the low temperature specific heat maximum on the abscissa axis. The analytical form of both equations is of course known but analytical solutions cannot be found.
At  temperature $t=t^*$, the specific heat shows a giant peak and the correlation functions a very sharp jump $G_{ss}$ from $\approx 1$ to $\approx -1$ , and $G_{ssb}$ from $\approx 1$ to $\approx 0$ (Fig.6). However, the effect is not abrupt and a very noticeable increase of the specific heat is observed in the finite range of $j_4 $, and for example for $j_4 = 0.2$ the value of the specific heat maximum is an order greater than for $ j_4=0$, for $ j_4=0.4$ by two orders, and for $ j_4=0.5$ by three orders, etc (Fig.5b). 

The crossover frustration-no frustration is not a phase transition, so the correlation length defined by
\begin{equation}
\frac{1}{\xi}= \ln[\frac{\exp(\beta j_4) \cosh (\beta j_t) + \sqrt{ \exp(2 \beta j_4) \sinh^2\beta j_t + \exp(-2 \beta j_4)}}
{\exp(\beta j_4) \cosh (\beta j_t) - \sqrt{ \exp(2 \beta j_4) \sinh^2\beta j_t + \exp(-2 \beta j_4)}}].
\end{equation}
is huge but finite at crossover tamperature ($t^*$). 
Fig.7 shows how inverse crorrelation length $1/\xi$ decreases as $j_4 \rightarrow j_4^* \approx 0.9$ at $t=t^* \approx 0.115415$ (for $J_y=-1, j_1=1.04$).

The effective inchain interaction $j_t$ depends on $J_y, j_1$ and temperature and does not depend on $j_4$, nevertheless a finite and positive  $j_4 > 0$  is a $\emph{sine qua non}$ condition for the occurrence of the giant specific heat (Figs.4 and 5) allowing to control the position of $t_{max}$. 

\begin{figure}
\label{Fig_8}
 \epsfxsize=10cm \epsfbox{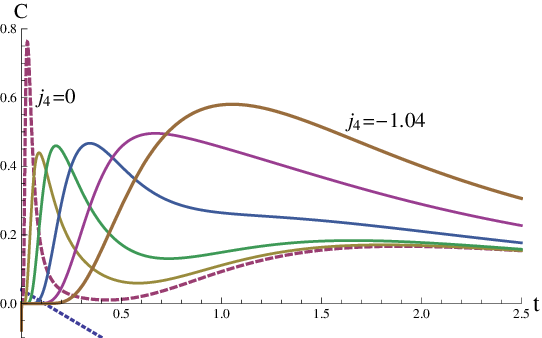}
 \caption{Temperature dependences of the specific heat for $J_y = -1, j_1 = 1.04$, and  $j_4=0$ (dashed line), $-0.1, -0.2, -0,4,-0.7, -1.04$, from left to right. Dotted line indicates the effective interaction $j_t$.}
 \end{figure}
 In papers Ref.~\onlinecite{Roj2, Roj3} the crossover (quasitransition) temperature is defined as a point in which the diagonal elements of the corresponding transfer matrix are equal to each other. For the present model it means the temperature in which the effective interaction parameter $j_t$ changes sign. As shown above, this is a necessary condition for the peculiarities in response functions to occur, however for the considered model this is not a sufficient condition.

\subsection{Renormalization Group Approach.}
In this subsection we apply to the model under consideration described by the Hamiltonian (4), the Linear Renormalization Group (decimation) method. Because the model is in fact one-dimensional such a transformation should reproduce the exact
result. Let us consider 4-spin block of the model (4)
\begin{equation}
H_4 = 2A + k_t(\Omega_1/2 + \Omega_2+ \Omega_3+ \Omega_4/2)+k_4(\Omega_1  \Omega_2+\Omega_2 \Omega_3 + \Omega_3  \Omega_4).
\end{equation}
Now  the factor ($-\beta$) has been absorbed in the Hamiltonian, $k_i = -\beta j_i$ and $A = \frac{1}{2} \ln( 4\cosh 2 k_1)$.

The RG transformation has the form

\begin{equation}
H'_4 = \ln(Tr_{\Omega_2, \Omega_3} e^{H_4}) = Z_0 + J_t (\Omega_1 +\Omega_4)+J_4 \Omega_1 \Omega_4,
\end{equation}
where

\begin{eqnarray}
Z_ 0 = \frac {1} {4}\ln \lambda_ 1  \lambda_ 2 \lambda_ 3^2, \quad J_t =\frac{1}{2}\ln \frac{\lambda_2}{\lambda_1}, \quad J_4=\frac{1}{4} \ln \frac {\lambda_1 \lambda_2}{\lambda_3^2}.
\end{eqnarray}
and
\begin{eqnarray}
\lambda_ 1 &=& e^{A - 3 k_t - k_4} (2 e^{2 k_t} + e^{4 k_t} + e^{4 k_4}), 
\lambda_ 2 = e^{A - k_t - k_4)}(1 + 2 e^{2 k_t} + e^{4 (k_t + k_4)}), \nonumber \\
\lambda_ 3 &=&  e^{A - 3 k_4} + e^{A + k_4} + e^{A - 2 k_t + k_4} + e^{A + 2 k_t + k_4}.
\end{eqnarray}
It is easy to check that the transformation (10-11) has only trivial fixed points at $T=0$ and $T=\infty$ and  the considered model does not exhibit any finite temperature phase transition. Evaluating numerically the RG transformation, one can find the free energy per site 
\begin{equation}
f= \sum_1^{\infty} \frac{Z_0^{(n)}}{3^n}
\end{equation}
and then the temperature dependence of the specific heat which reproduces exactly rigorous results as expected.

In conclusion in this section we focused on the Ising chains coupled by four-spin interaction $j_4 > 0$ with $J_y = -1$ and  $j_1 > -J_y$. The ground state of such a model is a fully ordered ferromagnetic in a chain, however the spins of the individual chains may be parallel (ferromagnetic) or antiparallel (antiferromagnetic) to each other. The positive four-spin interaction $j_4>0$ does not order the chains. At any finite temperature the long range order is destroyed, the correlation function $G_{ss} > 0$ (Fig.3c and 3d) and the system is unfrustrated up the the temperature $t = t^*(j_1, j_4)$. At this point the effective inchain interaction $j_t$ changes sign, the specific heat shows giant growth, the correlation function $G_{ss}$ a jump from $\approx +1$ to  $\approx -1$ and the system becomes frustrated. So, the giant specific heat indicates triggered by temperature a crossover from unfrustrated to frustrated state which in numerics could be confused with critical phase transition. As shown in Fig.8 the effect of the giant specific heat does not occur for a negative $j_4$ where the spins from $b$ subsystem are not fully frustrated at any temperature even though there is a point at which the effective interaction $j_t$ changes sign.
  
\section{2D Ising model with zig-zag interaction.}

In this section we consider a genuinely 2D Ising model with different two-spin interactions $j_p$ and $j_2$ in the $x$ and $y$ directions and zig-zag interaction $j_1$ (Fig.9) described by the Hamiltonian
\begin{equation}
H = -j_p \sum_{i,j} s_{i,j} s_{i+1,j} - j_1\sum_{i,j} (s_{i,j}+s_{i,j+1}) b_{i,j} - j_2 \sum_{{i,j}} s_{i,j} s_{i,j+1}.
\end{equation}

As before one can perform a strict summation over the spins $b_{i,j}$ degrees of freedom and obtain the Hamiltonian of the standard Ising model on the square lattice with temperature-dependent interchain interaction parameter $j_t$
\begin{equation}
H = - N A t - j_p \sum_{i,j} s_{i,j} s_{i+1,j} - j_t \sum_{i,j} s_{i,j} s_{i,j+1},
\end{equation}
where
\begin{equation}
j_t = j_2 + B t,
\end{equation}
$A$ and $B$ have been defined in (3), $t = k_B T/\mid j_p \mid$, and in numerical calculation we will always assume $j_p/k_B =1$. 

In order to find a critical temperature of the model (14) one can use the known exact Onsager's solution formula for  the anisotropic 2D Ising model 

\begin{equation}
\sinh \frac{2j_p}{t_c} \sinh \frac{2j_t}{t_c}=1.
\end{equation}
It is easy to check that due to the temperature dependence of $j_t$, for a given value of $j_2<0$ there is a range of $j_1$ values for which instead of one solution the equation (16) has three solutions. These solutions are presented in Fig.12 for $j_2=-1$ as thin solid lines. The same result, three successive phase transitions has already been found by using the Onsager equation (16) for decorated Ising lattice \cite{Itiro}.

\begin{figure}
\label{Fig_9}
 \epsfxsize=8cm \epsfbox{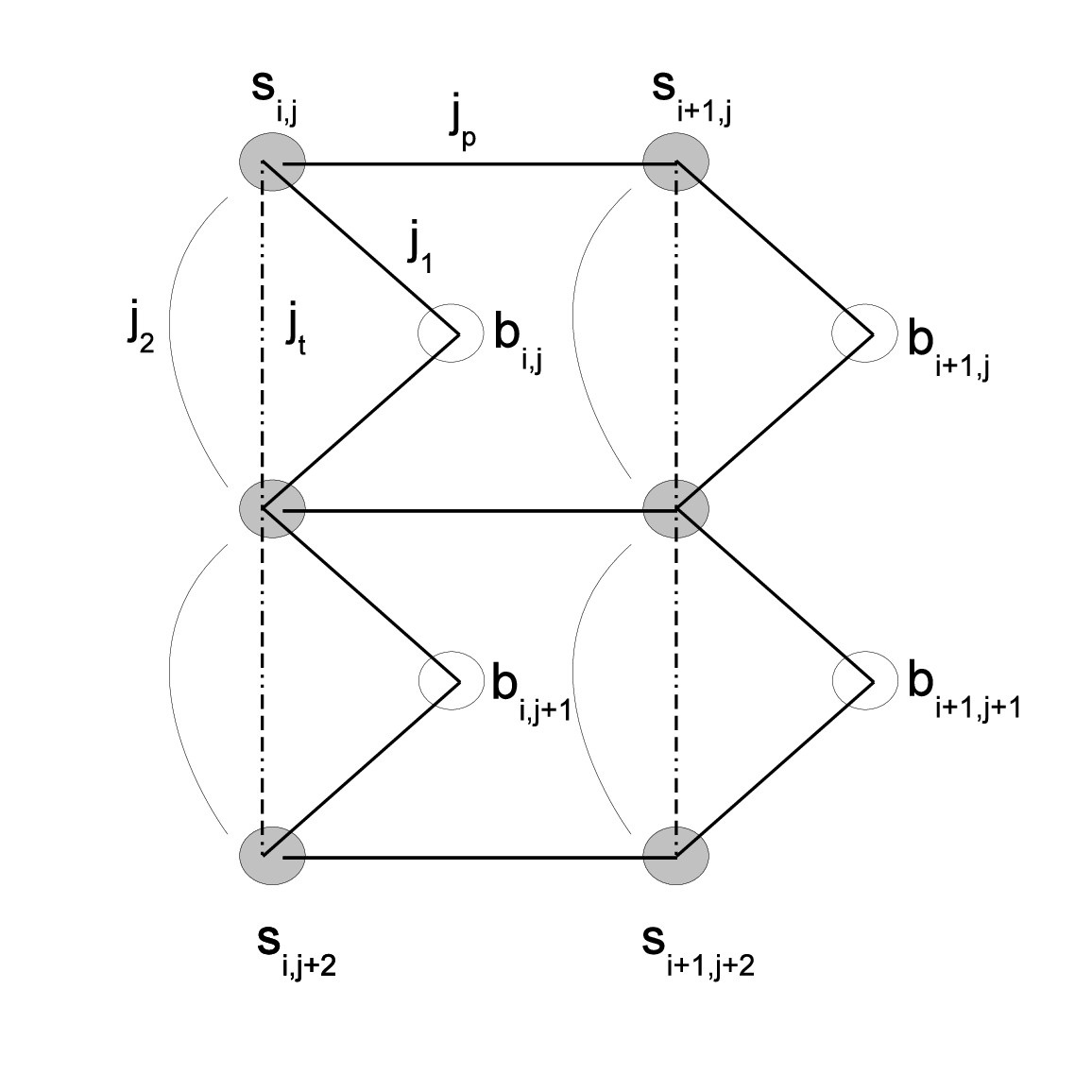}
 \caption{2D Ising model with zig-zag structure.}
 \end{figure}

To study the character of the singularity indicated by the equation (16) we use the linear perturbation renormalization group (LPRG \cite{JS1}) technique which for the Ising-type models starts with exact decimation of  one chain Hamiltonian 
\begin{equation}
H_0 =A N + k_p \sum_{i} s_{i,j} s_{i+1,j}.
\end{equation}
We should emphasized that as usual when using RG methods the factor $-\beta$ has been absorbed in the Hamiltonian, $k_i=-\beta j_i $,  and  $k_t = k_2 + B,   A = \frac{1}{2} \ln( 4\cosh 2 k_1),  B =  \frac{1}{2} \ln( \cosh 2 k_1)$.
Then, on the basis of it the interchain interaction
\begin{equation}
H_I =  k_t \sum_{i,j} s_{i,j} s_{i,j+1}
\end{equation}
is renormalized in a perturbative way. Denoting by $\sigma$ subsystem of decimated spins $s_{i,j}$ 
the renormalized Hamiltonian can be written in the form
\begin{equation}
H' = \ln(Tr_{\sigma} e^{H_0} ) + \ln\langle e^{H_I} \rangle_0
\end{equation}
where
\begin{equation}
\langle A \rangle_0 = \frac{Tr_{\sigma} A e^{H_0}}{Tr_{\sigma} e^{H_0}}.
\end{equation}

It was shown \cite{JS2} that LPRG method with the cluster presented in Fig.10  used to obtain the renormalized Hamiltonian leads to satisfactory results for the standard Ising model on the square lattice. For example in the third order cumulant expansion the LPRG gives the inverse critical temperature $k_p^c \approx 0.443$ which should be compared with the exact result $k_p^c \approx 0.4407$, and for the anisotropic case with inchain interaction $j_p=1$ and interchain $0.35 < j_t<1$ the deviation from the exact result is about $0.5 \%$.
 
\begin{figure}
\label{Fig_10}
 \epsfxsize=15cm \epsfbox{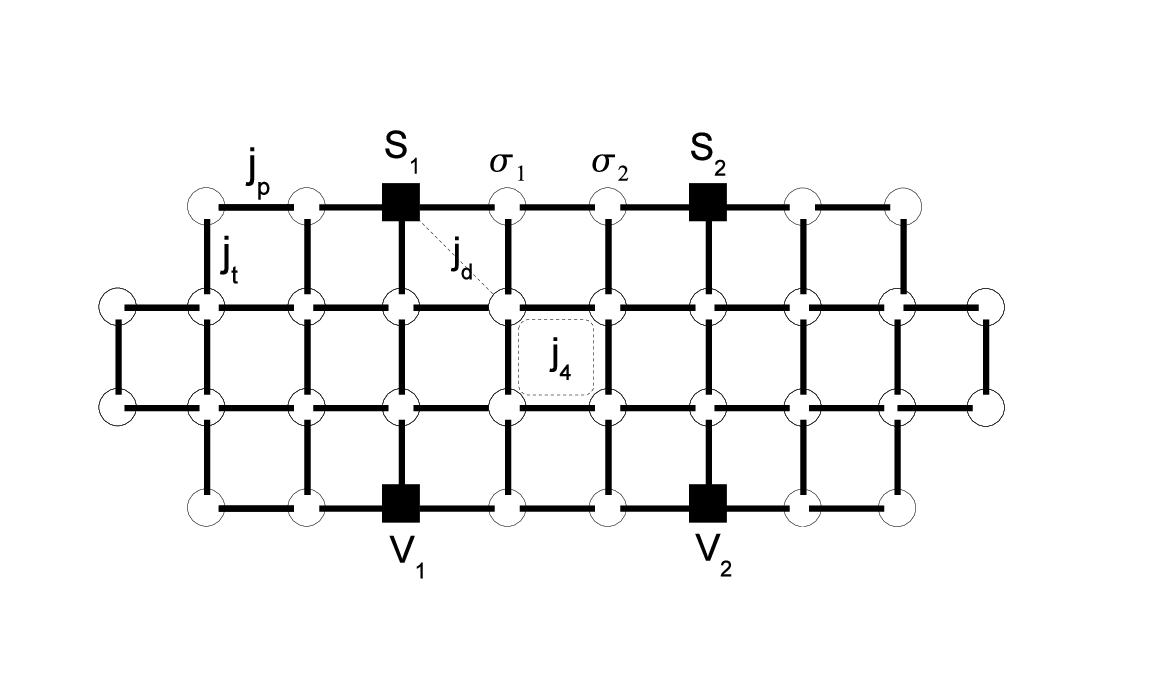}
 \caption{Cluster used to obtain the renormalized Hamiltonian of the model (13). Open circles denote decimated $\sigma$ and full squares survived $S$ quantum spins .} 
 \end{figure}

In the third order of the cumulant expansion, transformation (19) for the cluster presented in Fig.10 has the form
\begin{eqnarray}
H' &=& Z_0 + J_0 (S_1 S_2 + V_1 V_2)  +\nonumber \\
\langle H_I \rangle & +&\frac{1}{2} (\langle H_I^2 \rangle - \langle H_I \rangle^2) + \frac{1}{6} (\langle H_I^3 \rangle - 3 \langle H_I \rangle \langle H_I^2 \rangle + 2 \langle H_I \rangle^3),
\end{eqnarray}
where $S_i$ and $V_i$ spins  which survives in the RG procedure (Fig.10),
and
\begin{equation}
Z_0 =\frac{1}{2} \ln(10+3 e^{-4k_p}+3 e^{4k_p}), \quad J_0 =\frac{1}{2} \ln \frac{e^{2k_p}(3+e^{4k_p})}{1+3 e^{4k_p}}.
\end{equation}

The spin averages required to evaluate the transformation (21)  for a decimated row (first and fourth in Fig.10) read \cite{JS1}
\begin{equation}
\langle \sigma_1 \rangle = a_1 S_1 + a_2 S_2, \quad  \langle \sigma_2 \rangle = a_2 S_1 + a_1 S_2, \quad  \langle \sigma_1 \sigma_2 \rangle =a_1 + a_2 S_1 S_2,
\end{equation}
where
\begin{eqnarray}
a_1 &=& \frac{2(e^{8k_p}-1)}{(3+e^{4k_p})(2+3 e^{4k_p})}, \quad a_2 = \frac{(e^{4k_p}-1)^2}{(3+e^{4k_p})(2+3 e^{4k_p})}.
\end{eqnarray}
and for the removed rows (second and third in Fig.10) 
\begin{equation}
\langle \sigma _i  \rangle = 0, \quad \langle \sigma_i \sigma_{i+n} \rangle  = r, \quad\mbox{and} \quad  r = \tanh k_p.
\end{equation}
As usually the RG transformation generates new interaction which have to be included in the original Hamiltonian (14). In this case, these are two-spin diagonal interactions $j_d$ and four-spin $j_4$ .
\begin{equation}
 k_d \sum_{i,j} s_{i,j} s_{i+1,j+1} + k_4 \sum_{i,j} s_{i,j}  s_{i+1,j} s_{i,j+1} s_{i+1,j+1},
\end{equation}
($k_d=-\beta j_d, k_4=-\beta j_4$).
Now it is very easy to determine all the averages from equation (21) for example
\begin{equation}
\langle H_I \rangle = 3 a_2 r k_4 +(14 a_1+9 r)r k_4 (S_1 S_2+V_1 V_2),
\end{equation}
although the final formulas for renormalized interaction parameters are rather complicated.
Finally, the renormalized Hamiltonian takes the form
\begin{equation}
H' = Z + J_p (S_1 S_2+V_1 V_2) + J_t (S_1 V_1+S_2 V_2) + J_d (S_1 V_2+S_2 V_1) + J_4 S_1 S_2 V_1 V_2.
\end{equation} 
The recursion relations of the LPRG transformation read
\begin{equation}
J_{\alpha} = f_{\alpha}(k_p, k_t, k_d, k_4), \quad\mbox{with} \quad  k_t (j_2, j_1, t),
\end{equation} 
and the free energy can be calculated from the formula (12).
We should emphasize that RG approach relies on nonsystematic approximation and must be treated with caution. However, with respect to the existence of singularities, it seems plausible and confirmed by strict results for models for which such results are available. 

We have evaluated numerically the RG transformation from the original set of coupling parameters $j_{\alpha}$ to the set of renormalized parameters $J_{\alpha}$ and for $j_2 =-1$ and $j_1 \le 1$ ($j_t \le 0$ over the entire temperature range) 
have found two stable trivial fixed points at $J_{\alpha} = 0$ and $J_{\alpha} = \infty$ and the critical point corresponding to the continuous phase transition to an ordered phase.
The applied RG method does not allow to conclusively determine the character of the ordered phase  due to divergence of the interaction parameters however one can assume a partially frustrated antiferromagnetic system of the ferromagnetic chains ($j_p=1$).
For $1 < j_1 < j_1^c$ the RG transformation exhibits two nontrivial fixed points as shown in Fig.11. In the used LPRG approximation $j_1^c \approx 1.185$ and the exact result from the formula (16) is $j_1^c \approx 1.212$. With the increase of $j_1$ from $1$ the temperature $t_c^{(1)}$ and $t_c^{(2)}$ (Fig.12) approach each other and at $j_1 = j_1^c$, $t_c^{(1)}=t_c^{(2)} = t_b$. 
The relevant phase diagram in the plane $(t,j_t)$ is shown in Fig.12. 
For $ -j_2 < j_1 < j_1^c$ the system undergoes three continuous phase transitions from the disordered phase (D) to the ordered  ($'AF'$) then reentrant transition back to the disordered phase and finally to the ordered phase ($'FM'$). 
In Fig.13 the temperature dependence of the specific heat which diverges at three critical temperatures is presented.
As mentioned above on the basis of LPRG calculation we are not able to identify the type of spin order 
nevertheless it seems clear that phase ordered  ($'FM'$)
is fully ferromagnetic. The existence of the $'FM'$ phase, the only one ordered for $j_1 > j_1^c$, obviously has a connection with a change in the sign of $j_t$ from negative to positive.
However, the phase transition between disordered and $'FM'$ phases  is not a transition between frustrated - unfrustrated  phases and as seen in Fig.13  the temperature of this transition is not the same as the temperature of the sign change of the interaction parameter $j_t$.

\begin{figure}
\label{Fig_11}
 \epsfxsize=10cm \epsfbox{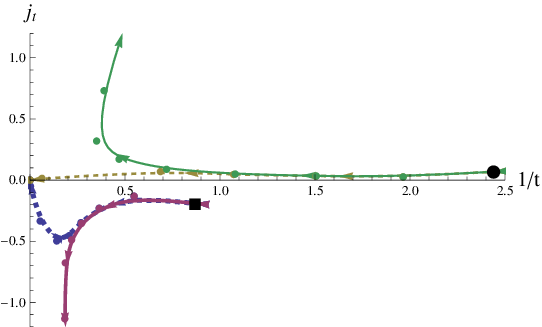}
 \caption{Illustration of the lines of flow in the plane $(j_t, 1/t)$ for the 2D Ising model (13) with $j_1=1.17$ and $j_2=-1$ from nontrivial antiferromagnetic fixed point (full square) and ferromagnetic fixed point (full circle) to disordered point (dashed lines) and to ordered ones (solid lines).} 
 \end{figure}

\begin{figure}
\label{Fig_12}
 \epsfxsize=12cm \epsfbox{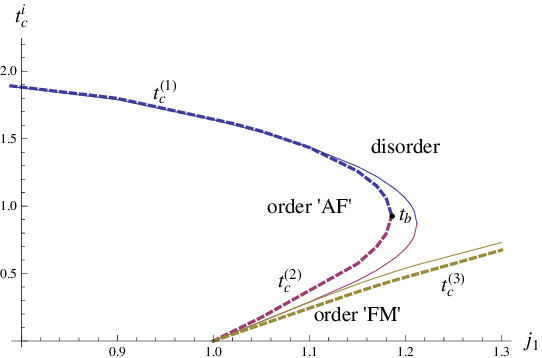}
 \caption{Phase diagram in the plane $(t_c, j_1)$ of the 2D Ising model with inchain $j_p=1$, interchain $j_2=-1$ and zig-zag $j_1$ interactions. The solid lines denote exact results from formula (16) and dashed lines from LPRG approximation.}
 \end{figure}

\begin{figure}
\label{Fig_13}
 \epsfxsize=12cm \epsfbox{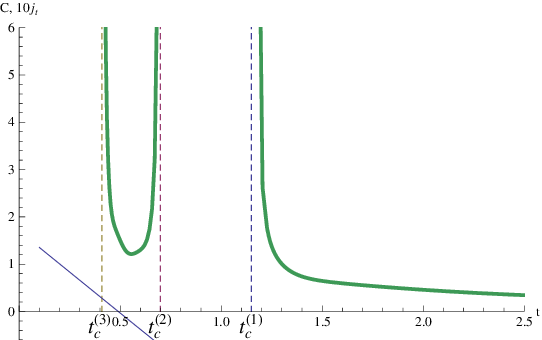}
 \caption{Temperature dependences of the specific heat  $C$ (solid lines) and effective interchain interaction $j_t$ (thin) for the 2D Ising model with $j_2 = -1$ and $j_1=1.117$ .}
 \end{figure}
 
\vspace*{2 cm}
 
\section{ Three phase transitions in $XXZ$ model.}
 
In several uranium and rare earth compounds the magnetic atoms form zig-zag chains. The mechanism responsible for the magnetic properties of such materials is a competition of interactions. In this paper we consider models whose structure makes it possible to sum over some degrees of freedom, reducing this competition of interactions to a single temperature-dependent interaction. The Ising model analyzed in the previous sections satisfactorily describes the basic features of phase transitions however, in real materials  transition scenario can, of course, be much more complex. Therefore, in this section we study the more general 2D quantum $XXZ$  model made of zig-zag chain  with easy axis along $z$ direction described by the Hamiltonian
 \begin{equation}
 H_{XXZ} = H_0 + H_{int}.
\end{equation}  
 where one chain part reads
\begin{equation}
H_0 = k_x \sum_{i,j} (s_{i,j}^x s_{i+1,j}^x + s_{i,j}^y s_{i+1,j}^y) +  k_z \sum_{i,j} s_{i,j}^z s_{i+1,j}^z, 
\end{equation} 
whereby for simplicity a zig-zag structure is adopted only for the $z$ spin component,
and interchain coupling is
 \begin{equation}
 H_{int} =  k_{x1} \sum_{i,j} (s_{i,j}^x s_{i,j+1}^x + s_{i,j}^y s_{i,j+1}^y)+  k_{z1} \sum_{i,j} s_{i,j}^z s_{i,j+1}^z + 
 k_1 \sum_{i,j}  (s_{i,j}^z+ s_{i,j+1}^z) b_{i,j}^z,
 \end{equation}
where $ \hat s_i =  (s_i^x, s_i^y, s_i^z)$ are Pauli  spin operators, $s_i^z = \pm 1$, the factor $-\beta$ has already absorbed in the Hamiltonian ($k_i = -\beta j_i$, $t = k_B T/\mid j_z \mid$ and  $|j_z| > |j_x|$.
In the isotropic case ($j_x=j_z, j_{x1}=j_{z1}$) and $j_1=0$ according to the Mermin-Wagner  theorem the long range order of the model (31-32) is constrained to zero temperatur. While the easy axis model studied in this paper exhibits finite temperature Ising type phase transition for all values of anisotropy cf., e.g., Ref.~\onlinecite{Far}. Since $j_1$ only changes the value of the effective interaction between chains, and does not change the symmetry of the problem cannot change the nature of the transition, if any takes place. 
 
 After rigorous summation over $b_i$ degrees of freedom the Hamiltonian takes the form
 
 \begin{equation}
 H_{int} = A N +  k_{x1} \sum_{i,j} (s_{i,j}^x s_{i,j+1}^x + s_{i,j}^y s_{i,j+1}^y)+  k_{t} \sum_{i,j} s_{i,j}^z s_{i,j+1}^z,
 \end{equation}
 where 
 \begin{equation}
 k_t = k_{z1} + B,  \qquad A = \frac{1}{2} \ln( 4\cosh 2 j_1) ,  \quad  B =  \frac{1}{2} \ln( \cosh 2 j_1).
 \end{equation}
 
To study the $XXZ$ model we use the LPRG method \cite{JS1} which starts with an approximate decimation of one dimensional system \cite{ST}. Then, similarly as for the Ising spins in the previous  section, the interchain interaction is renormalized in a perturbation way. We use the cluster presented in Fig.10 in third order of cumulant expansion. 
 As usual, the RG transformation from the original spins $\vec s_i$ to effective $\vec S_i$ is defined by
 \begin{equation}
 e^{H'(\vec{S})} = Tr_{\vec{s}} P(\vec S, \vec s) e^{H(\vec s)}
 \end{equation}
 and the weight operator $ P(\vec S, \vec s)$ is chosen in the linear form for four spin block (Fig.10)
 \begin{equation}
  P(\vec S, \vec s) = \frac{1}{4} (1 + \vec s_1 \vec S_1)(1 + \vec s_4 \vec S_2).
 \end{equation}
 In such a procedure the renormalized Hamiltonian of a single chain has the form
 \begin{eqnarray}
 H'_0 &=& \ln Z_0 , \quad Z_0 = L_0 +L_x  \sum_{i=1}^2 (S_{i}^x S_{i+1}^x + S_{i}^y S_{i+1}^y) +  J_{z0} ) \sum_{i=1}^2 S_{i}^z S_{i+1}^z, 
 \end{eqnarray}
 where
 \begin{equation}
L_0 = \frac{1}{4} Tr_{s} e^{H_0}, \quad L_x = \frac{1}{4} Tr_{s} s_1^x s_4^x e^{H_0}, \quad L_z = \frac{1}{4} Tr_{s} s_1^z s_4^z e^{H_0}
\end{equation} 
thus
\begin{equation}
H'_0 = G_0 + J_{x0} \sum_{i=1}^2 (S_{i}^x S_{i+1}^x + S_{i}^y S_{i+1}^y) +  J_{z0}  \sum_{i=1}^2 S_{i}^z S_{i+1}^z,
\end{equation}
where
 \begin{equation}
 G_0 = \frac{1}{4} \ln(\lambda_1 \lambda_2 \lambda_3^2), \quad 
 J_{x0} = \frac{1}{4} \ln \frac{\lambda_1}{\lambda_2}, \quad J_{z0} = \frac{1}{4} \ln \frac{\lambda_3^2}{\lambda_1 \lambda_2},
 \end{equation}
 and
\begin{equation}
\lambda_1 = L_0 + 2 L_x - L_z, \quad  \lambda_2 = L_0 - 2 L_x - L_z, \quad  \lambda_3 = L_0 + L_z.
\end{equation} 

 In this approach the LPRG generates two new interactions
 
 \begin{equation}
 H_{g} =  k_{x2} \sum_{i,j} (s_{i,j}^x s_{i+1,j+1}^x + s_{i,j}^y s_{i+1,j+1}^y)+  k_{z2} \sum_{i,j} s_{i,j}^z s_{i+1,j+j}^z.
 \end{equation} 
which must be included in the original Hamiltonian. Ultimately, the LPRG transformation from the original interaction parameters $k_i$ to the effective $J_i$  has the form of six recursion relations 

\begin{equation}
[k_x,  k_z,  k_{x1},  k_{x2},  k_{z2},  k_t(j_{z1},j_1,t)]  \quad  \rightarrow  \quad  (J_x, J_z, J_{x1}, J_{x2}, J_{z2}, J_t).
\end{equation}
 
To evaluate the LPRG transformation perturbatively using cumulant expansion (21) we have to know the averages of the spin components
\begin{equation}
\langle s_1^{\alpha} \rangle =  S_1^{\alpha}, \quad \langle s_4^{\alpha} \rangle = S_2^{\alpha}, \qquad \alpha =x,y,z,
\end{equation}
 and
 \begin{equation}
 \langle s_2^\alpha \rangle = (a_1^{\alpha} S_1^{\alpha} + a_2^ {\alpha} S_2^{\alpha}) Z_0^{-1}, \quad
  \langle s_3^\alpha \rangle = (a_2^{\alpha} S_1^{\alpha} + a_1^ {\alpha} S_2^{\alpha}) Z_0^{-1}, 
 \end{equation}
 with
 \begin{equation}
 a_1^{\alpha} = \frac{1}{4} Tr_{\vec s} s_2^\alpha  s_1^\alpha e^{H_0}, \quad a_2^{\alpha} =  \frac{1}{4} Tr_{\vec s} s_2^\alpha  s_4^\alpha e^{H_0}
 \end{equation}
 
\begin{equation}
\langle s_2^{\alpha} s_3^{\alpha} \rangle = (d_0^{\alpha} + d_{\alpha \alpha} S_1^{\alpha} S_2^{\alpha} + d_{\alpha \beta} S_1^{\alpha} S_2^{\beta} + d_{\alpha \gamma} S_1^{\alpha} S_2^{\gamma}) Z_0^{-1},
\end{equation} 
 where
 \begin{equation}
 d_0^{\alpha} =  \frac{1}{4} Tr_{\vec s} s_2^\alpha  s_3^\alpha e^{H_0}, \quad 
 d_{\alpha \alpha}=  \frac{1}{4} Tr_{\vec s} s_2^\alpha  s_3^\alpha s_1^\alpha  s_4^\alpha e^{H_0}, \quad  d_{\alpha \beta}=  \frac{1}{4} Tr_{\vec s} s_2^\alpha  s_3^\alpha s_1^\beta  s_4^\beta e^{H_0}.
 \end{equation}
Evaluating numerically the LPRG recursion relations, one can find the free energy per site according to the formula (12) and Hamiltonian flow in six-dimensional parameter space (effective coupling $k_t$  (34) is the function of $j_{z1}, j_1$ and temperature).
Henceforth, we will determine the values of two original parameters $j_z =1,  j_{z1} = -0.3$ and two generated by the LPRG transformation $j_{x2} = 0$ and $j_{z2} = 0$, so the role of only three variables $j_x,  j_{x1}$ and  $j_1$ will be studied. 

   In Fig.14 the specific heat for the system with  $j_{x1}=-0.2$ and $j_1 = 0.4$ as a function of temperature is presented for several values of $j_x =0, 0.6, 0.7$ and $0.8$. The thin dotted line denotes the temperature dependence of the effective interchain $s_{i,j}^z s_{i,j+1}^z$ interaction $j_t$. As seen for $j_x < 0.8$ the specific heat exhibits three very sharp peaks, corresponding to the continuous  phase transitions from disordered phase at high temperatures $t=t_c^{(1)}$  to the ordered phase $'AF'$, then reentrant phase transition to the disordered phase at $t=t_c^{(2)}$ and finally to the ordered phase $'FM'$ at $t=t_c^{(3)}$. For larger values of the inchain interaction $j_x$ the system undergoes only one phase transition to the phase $'FM'$. 
In Fig.15 the specific heat for fixed values of $j_x =0.7$ and $0.8$ for several values of $j_{x1}$ is presented.  
For $j_x=0.8$ the system shows only one phase transition for the whole range of values of $j_{x1}$     
 whereas for  $j_x=0.7$ there is only one phase transition for a sufficiently large $ | j_{x1} |   \ge   j_{x1}^t $  (for used set of parameters  $ j_{x1}^t \approx 0.34$).  
 
In this section, a model of spin chains with a magnetically easy axis along the $z$ axis ($|j_z| > |j_x|$) is considered. Thus, one could speculate that the effect of $x$ and $y$ terms in the Hamiltonian on the nature of phase transitions is negligible. As you can see this is not true because both inchain plain interactions $j_x $ and interchain $j_{x1} $ can significantly change the phase transitions scenario. In Fig. 16 the phase diagrams in the plane ($t, j_x$) and ($t,j_1$) with three possible phases are presented. For certain values of $j_1$ the line of the transition between phase disordered $D1$ and phase called $'AF'$ and the line  of the transition $D2$ - $'FM'$ coincide at point  $M_t$, which indicates a temperature-induced metamagnetic transition. 
 
 \begin{figure}
\label{Fig_14}
 \epsfxsize=8cm \epsfbox{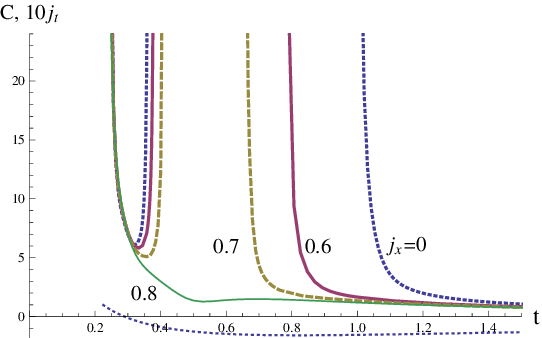}
 \caption{Temperature dependences of the specific heat of the 2D XXZ model with $j_1=0.4, j_{x1}=-0.2$, and $j_x=0$ ((dotted lines), $j_x=0.6$ (solid), $j_x=0.7$ (dashed), and $j_x=0.8$ (thin).}
 \end{figure}

 \begin{figure}
\label{Fig_15}
 \epsfxsize=16.5cm \epsfbox{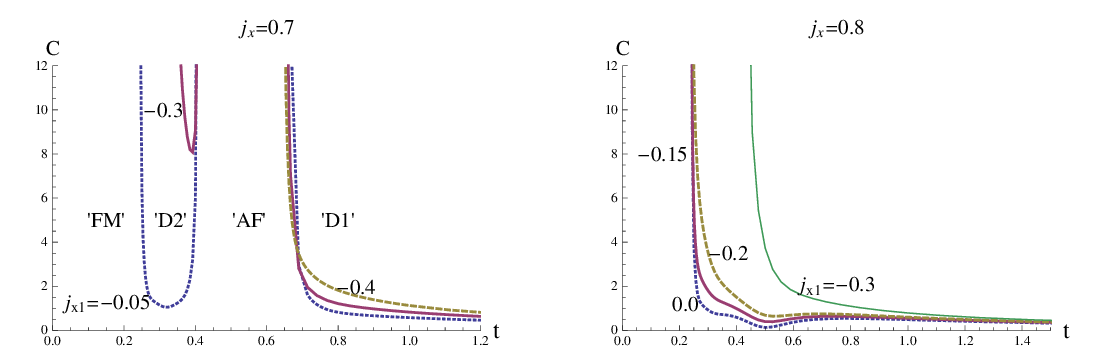}
 \caption{Temperature dependences of the specific heat 
 of the 2D XXZ model: left plot for $j_1=0.4, j_x=0.7$, and $j_{x1}=-0.05$(dotted line), $-0.2$ (solid), $-0.3$ (dashed), $-0.4$ (thin), and right plot for  $j_1=0.4, j_x=0.8$ and $j_{x1}=0.0$(dotted line), $-0.15$ (solid), $-0.2$ (dashed), $-0.3$ (thin). The thin dotted lines denote temperature-dependent effective interchain interaction $j_t$ (15).}
 \end{figure}
 
 \begin{figure}
\label{Fig_16}
 \epsfxsize=15.0cm \epsfbox{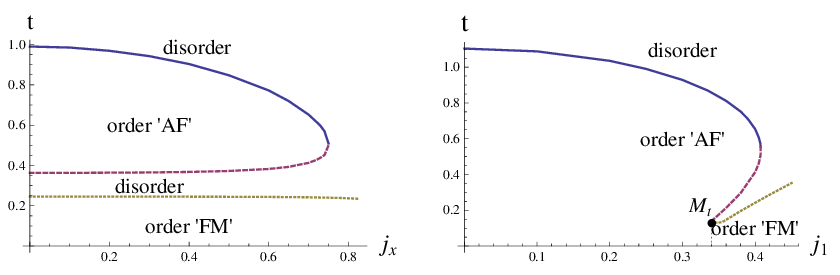}
 \caption{Phase diagrams of the 2D XXZ model in the plane $(t, j_x)$ for $j_{x1} = -0.2$ and $j_1 = 0.4$ (left plot) and in the plane  $(t, j_1)$ for $j_{x1} = -0.2$ and $j_x= 0.7$}
 \end{figure}
   
 \begin{figure}
\label{Fig_17}
 \epsfxsize=8cm \epsfbox{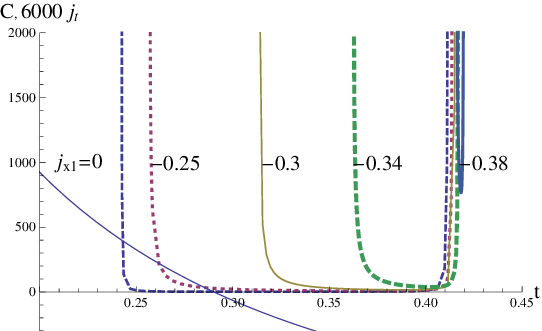}
 \caption{Temperature dependences of the specific heat of the 2D XXZ model for  $j_x=0.7$ and $j_{1}=0.34$, and $j_{x1}=0$ (thin line), $-0.12$ (dashed), $-0.13$ (dotted), and $-0.135$ (solid). The thin dotted lines denote temperature-dependent effective interchain interaction $j_t$ (14) multiplied by 1000 for visualization.}
 \end{figure}

In Fig.17 the range of disordered reentrant phase for a model with $j_x=0.7, j_1=0.34$, and several values of interchain  coupling $j_{x1}$ is shown. This range decreases with the increase of the parameter $|j_{x1}|$ and vanishes for a certain value of it. As seen for $j_{x1} > -0.25$ model the effective interchain interaction changes sign within the disordered phase while for $j_{x1} < -0.3$ in ordered phase $'FM'$ which character cannot be studied in the frame of the present method.
 
Finally, we should emphasize that the LPRG transformation for quantum systems is obtained by several approximations \cite{JS4}: that is, (i) the approximate decimation of one dimensional model (this procedure takes quantum effects within single block and neglects the effects of non commutativity of several blocks) which is not the case with an Ising like model, (ii) the approximate calculation of chain averages, (iii) the abbreviation of the cumulant expansion. The approximations (i) is a high temperature approximation and for sufficiently high temperature  leads to the values of the free energy of XY model \cite{JS4} very close to those found rigorously by Katsura \cite{Kat}. The approximation (iii) is justified for sufficiently weak interaction between chains.
In conclusion, as expected, the phase diagram of the quantum uniaxial $XXZ$ model resembles that of the Ising model with one difference. The point where the $'AF'$ and $'FM'$ phases meet (the metamagnetic transition) has been shifted from zero temperature for the Ising model to finite temperature for the $XXZ$ model. Given the approximate nature of the LPRG method
this result requires confirmation by complementary methods e.g. MC simulations.

 \section{Conclusion}
 
 We have considered three spin models made of zig-zag chains: (i) Ising chains coupled only by four-spin interaction (Fig.1), (ii) ferromagnetic Ising chains coupled by frustrated zig-zag two-spin interactions (Fig.9), and (iii)  coupled $XXZ$ chains with easy axis along "z" direction and zig-zag structure only for the $z$ spin component (31-32). Common features of these models are the frustration associated with the zig-zag structure and the possibility, by \emph {strictly} summing up with respect to some degrees of freedom, of replacing competing interactions with an effective temperature-dependent interaction. 
 
(i) Taking into account the character of the interchain coupling (only via four-spin interaction), the first model is in fact one dimensional and does not exhibit any finite temperature phase transition. Nevertheless it has been shown strictly that in the presence of the negative, antiferromagnetic inchain interaction $J_y < 0$ and positive interchain four-spin interaction $j_4 >0$, temperature triggers a frustration-no frustration crossover accompanied by gigantic specific heat. 
Huge specific heat and correlation length (close to zero but finite $1/\xi$) are found not only in the characteristic point of the crossover $j_4 = j_4^*$ but also in its surroundings (Fig.4 and Fig.7, respectively). Whereby this characteristic point is defined by such a value of $j_4$, for which the temperature at which the effective interaction $j_t$ (2) changes sign $t=t_p$, equals the temperature $t_{max}$, at which the spins in chain become unbound $(t_p = t_{max} = t^*)$. It was also rigorously shown that the giant specific heat does not occur for negative $j_4$ where spins from the $b_i$ subsystem are not fully frustrated at any temperature. The simple RG linear transformation reproduces  exactly the rigorous results.

(ii) In the genuine 2D Ising model (13) with antiferromagnetic direct interchain coupling $j_2 < 0$, the temperature-dependent effective interchain interaction $j_t = j_2+Bt, B =  \frac{1}{2} \ln( \cosh \frac{2 j_1}{t} )$ (15) obtained after strict summation with respect to the degrees of freedom of spins $b_i$ can result in three in place of the standard one phase transition. For $j_1 < -j_2$ the system exhibits only one phase transition from disordered $D$ phase to antiferromagnetically ordered ferromagnetic chains (for $j_p >0$) $'AF'$. However, for $- j_2 < j_1 < j_1^c$ the system shows sequence of three phase transitions with reentrant transition to the disordered phase  $D - 'AF' - D - 'FM'$. For $j_1 > j_1^c$ the system undergoes directly from disordered to $'FM'$ phase. Using the exact formula for the critical temperature of the 2D anisotropic Ising model (16) one can find exactly phase diagram in the plane $(t, j_1)$ (Fig.12) with one or three phase transitions. The three successive phase transitions were  already found for exactly solvable semi-decorated Ising model on the square lattice in  Ref.~\onlinecite{Itiro}.
The above mentioned phase diagram is reproduced satisfactory by the LPRG method (Fig.12).

(iii) The zig-zag structure has been found or is postulated in several uranium \cite{Aoki, Pikul, MP} and rear earth \cite{Amy} compounds. Particularly interesting are materials with competing interactions and possible frustrations such as $Sr R_2 O_4$ ($R = Ho, Dy$) \cite{Amy} or $UNi_{1-x} Ge_2$ \cite{MP, Pikul}. In this context, investigation of a minimal model showing the role of temperature-dependent interactions in spin systems, beyond the single-component Ising model, may be the way to explain the magnetic properties of such materials. In Sec IV simple easy axis $XXZ$ model with inchain ferromagnetic  and interchain antiferromagnetic interactions and with zig-zag structure adopted only for the $z$ spin component is considered. As with the Ising model, for a certain set of interaction parameters the model can undergo three phase transitions $'D1'-> 'AF' -> 'D2' -> 'FM'$ (Fig.15).
However, in this case both inchain plain interactions $j_x (s_{i,j}^x s^x_{i+1,j}+s_{i,j}^y s^y_{i+1,j})$ (in Fig.13 for $j_x \ge 0.8$) and interchain $j_{x1} (s_{i,j}^x s_{i,j+1}^x+s_{i,j}^y s_{i,j+1}^y)$ (in Fig.15 for $j_x=0.7$ and $j_{x1} \le -0.4$) can eliminate the $'D2'$ phase then the system shows only one transition. 
The decrease of the disordered phase range $'D2'$ with increasing $|j_{x1}|$ is depicted in Fig.17.
Although, as mentioned above, we are unable to determine the nature of the ordered phases within this method in  the case of the Ising model we can assume that phases $'AF'$ and $'FM'$ correspond to a simple arrangement of antiferromagnetically and ferromagnetically ordered spin chains, respectively. In the quantum $XXZ$ model such an assumption is not so obvious. Furthermore, while in the Ising model the lines of the $'D1' - 'AF'$ and $'D2' - 'FM'$ phase transitions coincide at t=0 (Fig.12), in the case of the $XXZ$ model this takes place at finite temperature at the point $M_t$ (Fig.16), which indicates a temperature-induced metamagnetic transition.


\end{document}